\PassOptionsToPackage{dvipsnames}{xcolor}
\documentclass[lettersize,journal]{IEEEtran}
\usepackage{amsmath,amsfonts}
\usepackage{algorithm}
\usepackage{array}
\usepackage[caption=false,font=normalsize,labelfont=sf,textfont=sf]{subfig}
\usepackage{textcomp}
\usepackage{stfloats}
\usepackage{url}
\usepackage{verbatim}
\usepackage{graphicx}
\def\BibTeX{{\rm B\kern-.05em{\sc i\kern-.025em b}\kern-.08em
				T\kern-.1667em\lower.7ex\hbox{E}\kern-.125emX}}
\usepackage{threeparttable} %
\IEEEoverridecommandlockouts
\usepackage{cite}

\usepackage{standalone}
\usepackage{orcidlink}

\usepackage{amssymb}
\usepackage{graphicx}
\usepackage{textcomp}
\usepackage{xcolor}
\usepackage{pgfplots}
\usepackage{bm}
\usepackage{tikz}

\usepackage{mathtools} %
\usepackage{comment}    %
\usepackage{nicefrac}

\usetikzlibrary{fit}
\usetikzlibrary{calc, shapes.gates.logic.US, shapes.gates.logic.IEC}
\usetikzlibrary{shapes.geometric, arrows.meta, calc, positioning, fit}
\usepgfplotslibrary{groupplots,dateplot}
\usetikzlibrary{positioning}
\usetikzlibrary{spy,backgrounds}
\usetikzlibrary{shapes.multipart}
\usetikzlibrary{positioning,calc}
\usepgfplotslibrary{fillbetween}
\usetikzlibrary{patterns}
\colorlet{mydarkblue}{blue!30!black}

\usepackage{multirow}  
\usepackage{placeins} %
\usepackage{soul}   %
\usepackage{algpseudocode}
\usepackage{footnote}
\makesavenoteenv{tabular}
\makesavenoteenv{table}

\pgfplotsset{compat=1.16}
\definecolor{darkblue}{rgb}{0,0.2706,0.541}
\definecolor{darkgreen}{rgb}{0.098,0.4784,0.5176}
\definecolor{lightgreen}{rgb}{0.3412,0.7411,0.7647}
\definecolor{orange}{rgb}{0.9255,0.4313,0}
\definecolor{darkred}{rgb}{0.7529,0,0}
\definecolor{black}{rgb}{0,0,0}
\def\BibTeX{{\rm B\kern-.05em{\sc i\kern-.025em b}\kern-.08em
    T\kern-.1667em\lower.7ex\hbox{E}\kern-.125emX}}

\definecolor{darkgray176}{RGB}{176,176,176}
\definecolor{goldenrod1911910}{RGB}{191,191,0}
\definecolor{lightgray204}{RGB}{204,204,204}
\definecolor{darkgray176}{RGB}{176,176,176}
\definecolor{darkviolet1910191}{RGB}{191,0,191}
\definecolor{green01270}{RGB}{0,127,0}
\definecolor{lightgray204}{RGB}{204,204,204}

\definecolor{myorange}{RGB}{236,110,0}

\definecolor{coloribx}{RGB}{0,69,138}
\definecolor{coloriby}{RGB}{192,0,0}
\definecolor{coloribt}{RGB}{25,122,132}
\definecolor{coloribs}{RGB}{10,90,90}

\newcommand\myrv[1]{\mathsf{\MakeUppercase{#1}}}
\newcommand\mylabel[1]{\mathsf{#1}}
\newcommand\myvec{\boldsymbol}%
\newcommand\mytuple{\bm}%
\newcommand\myset{\mathcal}
\newcommand{\breal}{b}

\newcommand{\hatl}{\hat{\ell}}
\newcommand{\labelcn}{\mathsf{c}}
\newcommand{\labelvn}{\mathsf{v}}
\newcommand{\labelch}{\mathsf{ch}}

\newcommand{\Z}{Z}

\newcommand{\iby}{y}
\newcommand{\ibY}{\myrv{Y}}
\newcommand{\ibyalp}{\mathcal{Y}}

\newcommand{\ibx}{x}
\newcommand{\ibX}{\myrv{X}}
\newcommand{\ibxalp}{\mathcal{X}}
\newcommand{\ibt}{t}
\newcommand{\ibT}{\myrv{T}}
\newcommand{\ibtalp}{\mathcal{T}}
\newcommand{\ibs}{s}
\newcommand{\ibS}{\myrv{S}}

\newcommand{\quant}{Q}

\newcommand{\p}{p} %

\newcommand{\matH}{\mathbf{H}}
\newcommand{\matHb}{\matH_\mylabel{b}}

\newcommand{\idxi}{i}

\newcommand{\idxj}{j}

\newcommand{\Hbij}{\mathsf{H}^{\idxi\idxj}_\mylabel{b}}

\newcommand{\opcol}{\operatorname{col}}
\newcommand{\oprow}{\operatorname{row}}
\newcommand{\opsgn}{\operatorname{sgn}}
\newcommand{\B}{\mathsf{B}}
\newcommand{\vecb}{\myvec{b}}

\newcommand{\dc}{\mathsf{d}^\labelcn}
\newcommand{\dv}{\mathsf{d}^\labelvn}
\newcommand{\setN}{\mathcal{N}}
\newcommand{\n}{n}
\newcommand{\opNv}{\operatorname{N}^\labelvn}
\newcommand{\opNc}{\operatorname{N}^\labelcn}
\newcommand{\tilden}{\tilde{n}}

\newcommand{\setA}{\mathcal{A}}

\newcommand{\setUv}{\mathcal{U}^\labelvn}
\newcommand{\setUc}{\mathcal{U}^\labelcn}

\newcommand{\tuplesetU}{\mytuple{\myset{U}}}
\newcommand{\itermax}{\iota_\mylabel{max}}
\newcommand{\iteravg}{\iota_\mylabel{avg}}

\newcommand{\tch}{t^\labelch}
\newcommand{\Tch}{\myrv{T}^\labelch}

\newcommand{\wch}{w^\labelch}
\newcommand{\Tc}{\myrv{T}^\labelcn}
\newcommand{\tc}{t^\labelcn}

\newcommand{\Sc}{\myrv{S}^\labelcn}
\newcommand{\scmy}{s^\labelcn}

\newcommand{\Tv}{\myrv{T}^\labelvn}
\newcommand{\tv}{t^\labelvn}

\newcommand{\barTc}{\bar{\myrv{T}}^\labelcn}

\newcommand{\bartcalp}{\bar{\mathcal{T}}^\labelcn}

\newcommand{\barSc}{\bar{\myrv{S}}^\labelcn}
\newcommand{\barsc}{\bar{s}^\labelcn}

\newcommand{\x}{x} %
\newcommand{\X}{\myrv{X}}

\newcommand{\barx}{\bar{\x}}
\newcommand{\barX}{\bar{\X}}

\newcommand{\Eb}{E_b}
\newcommand{\No}{N_0}

\newcommand{\EbNo}{\Eb/\No}
\newcommand{\I}{\operatorname{I}}

\newcommand{\lv}{\ell^\labelvn}

\newcommand{\lc}{\ell^\labelcn}
\newcommand{\lcminus}{\ell^{\labelcn,-}}
\newcommand{\Lc}{\myrv{L}^\labelcn}

\newcommand{\ulc}{\underline{\ell}^\labelcn}

\newcommand{\barlc}{\bar{\ell}^\labelcn}
\newcommand{\barLc}{\bar{\myrv{L}}^\labelcn}

\newcommand{\oprnd}{\operatorname{rnd}}
\newcommand{\phic}{\phi^\labelcn}

\newcommand{\phich}{\phi^\labelch}

\newcommand{\TP}{\operatorname{TP}}
\newcommand{\Nd}{N_\mathsf{d}}

\newcommand{\Nlayersr}{N_{\mathsf{layers}}^{(r)}}

\newcommand{\Agr}{A_\mathsf{g}^{(r)}}

\newcommand{\AEmy}{\operatorname{AE}}
\newcommand{\Zp}{Z_\mylabel{p}}

\usepackage{hyperref}
\usepackage[capitalise]{cleveref}
\crefformat{equation}{(#2#1#3)}
\crefrangeformat{equation}{(#3#1#4--#5#2#6)}
\crefmultiformat{equation}{(#2#1#3)}{ and~(#2#1#3)}{, (#2#1#3)}{, and~(#2#1#3)}

\crefname{algorithm}{Alg.\!}{Algs.\!} %
\Crefname{algorithm}{Alg.\!}{Algs.\!} %

\crefname{line}{Ln.\!}{Lns.} %
\Crefname{line}{Ln.\!}{Lns.} %

\IEEEaftertitletext{\vspace{-1\baselineskip}}
\def\mymodestandalone{buildmissing}

\begin{document}
\bstctlcite{IEEEexample:BSTcontrol} %
\title{
Memory-Assisted Quantized LDPC Decoding
\vspace{-0.2cm}
}
\author{Philipp Mohr\textsuperscript{\orcidlink{0000-0003-4350-9969}},
	\IEEEmembership{Graduate Student Member, IEEE}, and Gerhard Bauch, \IEEEmembership{Fellow, IEEE}
	\thanks{Manuscript received August 21, 2024; revised January 21, 2025. Philipp Mohr and Gerhard Bauch are with the Institute of Communications, Hamburg University of Technology, Hamburg, 21073, Germany. E-mail: \{philipp.mohr; bauch\}@tuhh.de.}
}

\markboth{IEEE COMMUNICATIONS LETTERS}%
{How to Use the IEEEtran \LaTeX \ Templates}
\maketitle

\begin{abstract}
We enhance coarsely quantized LDPC decoding by reusing computed check node messages from previous iterations. Typically, variable and check nodes update and replace old messages every iteration. We show that, under coarse quantization, discarding old messages entails a significant loss of mutual information. The loss is avoided with additional memory, improving performance by up to 0.23\,dB. We optimize quantization with a modified information bottleneck algorithm that considers the statistics of old messages. A simple merge operation reduces memory requirements. Depending on channel conditions and code rate, memory assistance enables up to 32\,\% better area efficiency for 2-bit decoding.
\end{abstract}
\FloatBarrier
\vspace{-0.1cm}
\section{Introduction}
\IEEEPARstart{E}{fficient} and reliable decoding of low-density parity-check (LDPC) codes is vital in modern technologies with high data rate requirements, such as 5G\cite{3gpp18}.
Particularly the exchange of messages in iterative message passing decoding algorithms like belief propagation demands significant complexity\cite{gal62}.
To overcome this bottleneck many works focus on reducing the bit width of the exchanged messages in these algorithms through quantization operations, see e.g. \cite{chen05,kang22,wang22,ren24,lew18,stark20,mohr22aware,mohr2024region}.

The quantized messages represent reliability levels that encode probability information exchanged between variable nodes~(VNs) and check nodes~(CNs). 
The choice of reliability levels is crucial for excellent decoding performance with low-resolution messages~\cite{mohr2024region}.
Information optimum reliability levels can be found with the information bottleneck (IB) method which is a clustering framework, that enables the design of compression operations for maximizing preserved relevant mutual information\cite{chechik2002extracting,steiner2021distributed, lew18,stark20,mohr22aware,mohr2024region}.
Relevant mutual information measures the average amount of information between the transmitted code bits and exchanged decoding messages.

Typically, calculated messages from a previous iteration are replaced by updated messages from the current iteration~\cite{chen05,lew18,stark20,kang22,wang22,mohr22aware,mohr2024region,ren24}.
One might question whether discarding previously computed and exchanged messages wastes valuable information.
Indeed, under coarse quantization, this work confirms that preserving old messages of the previous iteration can significantly improve the decoding performance.
For the design of a memory-assisted decoder we modify the sequential IB algorithm from \cite{lew18} to be aware of the messages retained in memory.
This algorithm is specifically suited for the design of deterministic compression mappings realized with symmetric thresholds.
It has significantly reduced computational costs compared to more general solutions\cite{steiner2021distributed}.

We combine the memory-assisted decoder structure with our recently proposed region-specific CN-aware quantizer design~\cite{mohr2024region}.
Region-specific quantization allows individual alphabets of reliability levels for subsets of exchanged messages particularly improving low-resolution decoding of highly irregular 5G-LDPC codes.
A CN-aware quantizer design for the VN extends the optimization scope to maximize preserved relevant information at the output of the subsequent CN update~\cite{mohr22aware}.
The combination of this work and \cite{mohr2024region} yields up to 0.68\,dB gain w.r.t. 2-bit decoding without those techniques.

\section{Preliminaries on LDPC Decoding with Mutual Information Maximizing Quantization}\label{sec:preliminaries}
We consider LDPC codes defined through a base matrix~$\matHb$ with entries $\Hbij{\in}\{-1,\ldots,\Z\}$\cite{3gpp18}.
The base matrix ~$\matHb$ can be represented by a Tanner graph.
Each column~$\idxj$ becomes a variable node~(VN) and each row~$\idxi$ becomes a check node~(CN).
The non-negative entries $\Hbij$ are edges between VNs and CNs.
The node degree, i.e., the number of edges connected to a node, is $\dv_\idxj$ for a VN and $\dc_\idxi$ for a CN.
Lifting replaces each edge with $\Z$ edges, which are subjected to a $\Hbij$-cyclic permutation.
The lifted graph can be equivalently represented by a lifted parity check matrix $\matH$.

The encoder maps the information bits $\myvec{u}$ to the code bits $\vecb$ such that $\matH\vecb{=}\myvec{0}$.
In 5G, puncturing causes only a subset of $\vecb$ to be transmitted to increase the code rate. We consider a memoryless AWGN channel.

All code bits related to a base column $\idxj$ are modeled with the random variables $\B_\idxj$ whose realizations are $\breal_\idxj{\in}\{0,1\}$.
The decoder observes $\wch$-bit channel messages modeled with $\Tch_\idxj,\tch_\idxj{\in}\mathcal{T}_{\wch}$ using a sign-magnitude alphabet $\mathcal{T}_\zeta{=}\{{-}2^{\zeta-1}, \ldots,{-}1,{}1,\ldots,{}2^{\zeta-1}\}$.
Message passing decoding computes and exchanges messages between VNs and CNs to aggregate soft information for error correction from the parity check constraints.
Each edge of the base graph enumerated by $\n{\in}\setN{=}\{1,\ldots,\sum_\idxj \dv_\idxj\}$ contains two memory locations, one for messages from VNs and one for messages from CNs.
A memory location stores $\Z$ messages after lifting the graph.
The sets $\setUv{\subseteq}\setN$ and $\setUc{\subseteq}\setN$ specify target memory locations for VN and CN updates, respectively.
The decoding schedule defines the order in which memory locations are updated as $\tuplesetU=(\setUv_0,\setUc_1,\setUv_2,\setUc_3,\ldots)$
followed by a final hard decision update using the latest updated CN messages. 
We model exchanged  $w$-bit VN and CN messages with discrete random variables $\Tv_\n, \tv_\n{\in} \mathcal{T}_{w}$ and $\Tc_\n, \tc_\n{\in} \mathcal{T}_{w}$, respectively.

Updating a VN memory location $\n{\in}\setUv$ yields
\begin{align}
	\lv_\n &= \phich_{\opcol(n)}(\tch_{\opcol(n)})+\sum_{\mathclap{m\in \tilden}} \phic_{\alpha(m)}(\tc_m)%
	\label{eq:vnu}
\end{align}
with extrinsic CN locations $\tilden{=}\opNv(\opcol(n)){\setminus}\{n\}$ where $\opNv(\idxj){=}\{m{\in}\setN{:}\opcol(m){=}\idxj\}$. %
The reconstructions functions $\phich_\idxj(t){=}\oprnd_\kappa(L(\B_\idxj|\Tch_\idxj{=}t))$ and $\phic_a(t){=}\oprnd_\kappa(L(\barTc_a{=}t|\barX_a))$ map incoming messages to integers corresponding to the underlying LLR levels.
The LLR functions are defined as $L(\myrv{T}{=}t|\myrv{X}){=}\allowbreak\log \frac{\p(\myrv{T}{=}t|\myrv{X}{=}0)}{\p(\myrv{T}{=}t|\myrv{X}{=}1)}$ and $L(\myrv{X}|\myrv{T}{=}t){=}\allowbreak L(\myrv{T}{=}t|\myrv{X}) {+} \log \frac{\p(\myrv{X}{=}0)}{\p(\myrv{X}{=}1)}$.
Further, $\oprnd_\kappa(\ell){=}\opsgn(\ell)\lfloor |\ell|/\kappa {+} 0.5\rfloor$ scales and rounds $\ell$ to integers with scaling parameter $\kappa$ \cite{mohr2024region}.
An implementation can use $\phic_a(t){=}\opsgn(t)\phic_a(|t|)$ for reducing complexity.

For the CN messages we allow different sets of LLR levels $\phic_a,a{\in}\setA$.
Each memory location $n$ is assigned to a region $a$ with $\alpha\,{:}\,\setN{\to}\setA$.
This work considers a row-alignment $\alpha(\n){=}\oprow(n)$ or 
matrix-alignment $\alpha(n){=}1$.
We use aligned random variables $\barTc_a{=}\allowdisplaybreaks E_n\{\Tc_{n}|\alpha(n){=}a\}$ and $\barX_a{=}\allowdisplaybreaks E_n\{\myrv{B}_{\operatorname{col}(n)}|\alpha(n){=}a\}$\cite{mohr2024region}.

The hard decision is modeled with $\hat{b}_{j}{=}(1{-}\operatorname{sgn}(\hat{\ell}_j))/2$ obtained from the a-posteriori probability (APP) LLR $\hatl_{\opcol(n)}=\allowdisplaybreaks\lv_n {+} \phic_{\alpha(n)}(\tc_{n})$.
Row-layered decoder structures, such as the one in \cref{fig:hardware}, typically use the APP LLR for computing \cref{eq:vnu} efficiently as $\lv_n{=}\hatl_{\opcol(\n)}{-}\phic_{\alpha(n)}(\tc_{n})$.
Those decoders initialize the APP LLR with $\hat{\ell}_\idxj{=} \phich_\idxj(\tch_\idxj)$.
Hence, the reconstruction of the channel message must be done only once for all iterations.

Updating a CN memory location $\n{\in} \setUc$ using the min-sum approximation\cite{chen05} yields
\begin{align}
	\tc_\n=Q_{\alpha(\n)}(\lc_\n) \text{ using } \lc_\n &= \prod_{\mathclap{m\in \tilde{n}}}\opsgn(\lv_{m}) \min_{\mathclap{m\in\tilde{n}}} |\lv_{m}|
	\label{eq:cnu}
\end{align}
with extrinsic VN locations~$\tilden{=}\opNc(\oprow(n)){\setminus}\{n\}$ and the CN neighborhood set $\opNc(i){=}\{m{\in}\setN{:}\opcol(m){=}i\}$.
Considering the quantizer after the CN update enables a CN-aware quantizer design as proposed in~\cite{mohr2024region}.
For all $a{\in}\{\alpha(\n){:}\n{\in}\setUc\}$, the quantizer design objective is to maximize mutual information $\max_{Q_a} \I(\barX_a;\barTc_a)$ preserved by any CN message with index~$\n$ where $\alpha(\n){=}a$.
The optimization of $\quant_a$ can be performed with the sequential IB algorithm\cite{lew18}, also described in \cref{sec:sideinfoib}. This algorithm requires the average joint probability mass function $\p(\barx_a,\barlc_a){=}E_{\n|a}\{p(\X_\n{=}\barx_a,\Lc_n{=}\barlc_a)\}$.

With the same behavior regarding $\tc_\n$ in \cref{eq:cnu}, the threshold quantizer $\quant_a$ can be moved before the CN update, reducing complexity of the $\min$ operation.

\section{Design with Memory-Assisted Reconstruction}\label{sec:designmemassistance}
\begin{figure}[t]
	\centering
	\includestandalone[mode=\mymodestandalone]{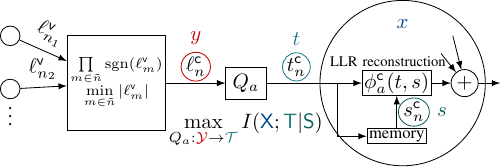} 
	\vspace{-0.2cm}
	\caption{Quantizer design with memory-assisted reconstruction.}
\label{fig:memawaredesign}
\end{figure}
Lowering the bit width $w$ of the exchanged messages simplifies decoder complexity for reconstruction, quantization, shifting networks and the min-sum CN update. 
However, very coarse quantization $w{<}4$ can significantly degrade the decoding performance\cite{mohr2024region}.
This section proposes to overcome most of the degradation by preserving the previous iteration's CN message $\tc_\n$ as $\scmy_\n$ to serve as side information in the VN reconstruction, as shown in \cref{fig:memawaredesign}.

The message $\scmy_\n$ conveys information w.r.t. $\x_\n$ such that $I(\X_\n; \Sc_n){>}0$.
The optimization of the quantizer shall optimize preservation of additional mutual information $I(\X_\n;\Tc_\n|\Sc_\n)$.
The variable node update now takes into account $\scmy_n$ as
\begin{align}
	\lv_\n &= \phich_{\opcol(n)}(\tch_{\opcol(n)})+\sum_{m\in \tilden} \phic_{\alpha(m)}(\tc_{m},\scmy_m)
	\label{eq:vnumemaware}
	\vspace{-0.3cm}
\end{align}
For the design of $\quant_{a}:\barLc_a{\to}\bartcalp_a$ in \cref{eq:cnu} we measure the joint distribution $\p(\barx_a,\barlc_a, \barsc_a)$ which considers correlations between $\tc_\n$ and $\scmy_\n$. 
Therefore, we generate a large set of decoding messages $\x_n,\lc_n, \scmy_n$ for a specific design-$\EbNo$.
In the next \cref{sec:sideinfoib} we introduce a side-information aware IB algorithm which is used for optimization of the quantization thresholds. 
We define $\barX_a$, $\barLc_a$, $\barTc_a$, and  $\barSc_a$ as the relevant, observed, compressed and side-information variable, $\ibX$, $\ibY$, $\ibT$, and $\ibS$, respectively.
The optimization aims for $\max_{p(\ibt|\iby)}\I(\ibX;\ibT|\ibS)$.
We remark that the alphabet $\ibyalp$ is only approximately LLR-sorted as defined in \cref{eq:llrsorting} as a result of the min-sum approximation. 

\section{Information Bottleneck Algorithm with Side-Information Awareness}\label{sec:sideinfoib}
Typically, an IB setup is defined by a relevant, observed and compressed random discrete variable $\ibX,\ibx{\in}\ibxalp{=}\{0,1\}$, $\ibY,\iby{\in}\ibyalp{\in}\{1,\ldots,|\ibyalp|\}$ and $\mathsf{T},t{\in}\mathcal{T}{=\{1,\ldots,|\mathcal{T}|\}}$ that form a Markov chain $\ibX{\to}\ibY{\to}\ibT$.
The IB method is a generic clustering framework for designing compression operations $\p(\ibt|\iby)$ with optimization objective $\max_{\p(\ibt|\iby)}\I(\ibX;\ibT)-\beta ^{-1} I(\ibY;\ibT)$. 
The choice of $\beta\ge0$ allows to trade preservation of relevant information $\I(\ibX;\ibT)$ for compression.
Of very high practical interest is the case where $\beta{\to}\infty$ because it can be achieved with a deterministic mapping through threshold quantization
\begin{align}
	\ibt = \quant(\iby) = \ibtalp[k] \quad \tau_k{\leq} y {<} \tau_{k+1}, 0{\le}k{<}|\ibtalp|
\end{align}
with outer thresholds $\tau_0{=}0$ and $\tau_{|\ibtalp|}{=}|\ibyalp|$, and $\ibtalp[k]$ identifying the $k$th element of the ordered set $\ibtalp$.
The mapping with thresholds can be information-optimum if
\cite{kurkoski_quantization_2014}
\begin{align}
	L(\ibx|\iby{=}1){\le}L(\ibx|\iby{=}2){\le}{\ldots}{\le}L(\ibx|\iby{=}|\ibyalp|).\label{eq:llrsorting}
\end{align}

This section extends the conventional setup with a fourth variable $\ibS$ which provides side-information about $\ibX$\cite{steiner2021distributed}. %
Prior works\cite{chechik2002extracting,steiner2021distributed} also consider a setup with side information, however, those works do not explicitly provide a low-complexity solution for optimizing a threshold quantizer in the context of LDPC decoding.
We propose \cref{alg:sibalgorithm} which is a modified variant of the sequential IB algorithm from \cite{lew18} taking into account the side information~$\ibS$. 
The algorithm exploits \cref{eq:llrsorting} to sequentially  optimize initial random boundaries $\tau_k$  defining the target clusters $\mathcal{Y}_t{=}\{y{:}\tau_t{\le}y{<}\tau_{t+1}\}{\subset}\ibyalp$. To avoid local optima, we run 500 different initializations in parallel. We enforce symmetric thresholds $\tau_k{=}|\ibtalp|{-}\tau_{|\mathcal{T}|-k}$, reducing the number of design and implementation parameters. 
\renewcommand{\algorithmicrequire}{\textbf{Input:}}
\renewcommand{\algorithmicensure}{\textbf{Output:}}
\algrenewcommand\algorithmicindent{0.7em}%
\begin{algorithm}[t]
	\caption{A sequential IB algorithm from \cite{lew18} considering side information $s$ in the merger cost computation (\cref{line:mergercostcomputation}).}\label{alg:sibalgorithm}
	\begin{algorithmic}[1]
		\footnotesize
		\Require $p(x,y,s),|\mathcal{T}|$
		\Ensure $p(t|y)$, $p(x,t,s)$
		\State Create random symmetric clustering $p(t|y)$;
		\State Compute \vspace{-0.5cm}\begin{align}
			p(x,t,s)=\sum_y p(t|y)p(x,y,s)\label{eq:updatepxts}
		\end{align}
		\vspace{-0.35cm}
		\State Extend clusters $\mathcal{Y}_{1},\ldots,\mathcal{Y}_{|\mathcal{T}|}$ with empty singleton clusters $\mathcal{Y}_{|\mathcal{T}|{+}1}$ and $\mathcal{Y}_{|\mathcal{T}|{+}2}$;
		\Repeat
		\State Save $p(t|y)$ as $p^{\mathrm{old}}(t|y)$;
		\For{$t \in \{1,\ldots,|\mathcal{T}|/2{-}1\}$}
		\For{$(t_1,t_2) \in \{\textcolor{red}{(t,t{+}1)},\textcolor{blue}{(t{+}1,t)}\}$}		
		\Repeat%
		\If{\textcolor{red}{$t_1==t$}}
		\State \textcolor{red}{$y$ is rightmost element in cluster $\mathcal{Y}_{t_1}$};
		\ElsIf{\textcolor{blue}{$t_1==t{+}1$}}
		\State \textcolor{blue}{$y$ is leftmost element in cluster $\mathcal{Y}_{t_1}$};
		\EndIf
		\State\label{line:movedy}Move $y$ into $\mathcal{Y}_{|\mathcal{T}|{+}1}$, and $y'{=}|\mathcal{Y}|{+}1{-}y$ into $\mathcal{Y}_{|\mathcal{T}|{+}2}$;\label{lst:extendedclusters}
		\State Update $p(x,t,s)$ with \cref{eq:updatepxts} where $p(t|y)$ considers \cref{line:movedy};
		\State Compute optimal cluster\label{lst:optcluster} using \cref{eq:symmergercosts}:\label{line:mergercostcomputation}\vspace{-0.15cm}
		\begin{align}
			k^*=\arg\min_{k\in\{1,2\}} C_\mylabel{sym}(y,k)
		\end{align}
		\includestandalone[mode=\mymodestandalone]{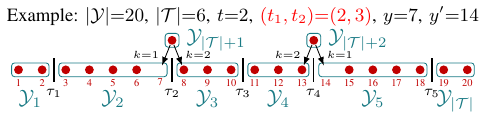}
		\State Merge $\mathcal{Y}_{|\mathcal{T}|{+}1}$ into $\mathcal{Y}_{t_{k}}$, and $\mathcal{Y}_{|\mathcal{T}|{+}2}$ into $\mathcal{Y}_{|\mathcal{T}|{+}1{-}t_{k^*}}$;
		\State Update $p(x,t,s)$ and $p(t|y)$;
		\Until{$t_k==t_1$}
		\EndFor
		\EndFor
		\Until{$p(t|y)==p^{\mathrm{old}}(t|y)$}
		\State \Return $p(t|y)$, $p(x,t,s)$
	\end{algorithmic}
\end{algorithm}

		\vspace{-0.4cm}
		\subsection{Merging Costs with Side Information}
		In line \ref{lst:extendedclusters}, the element $y$ and counterpart element $y'$ are moved into singleton clusters $\mathcal{Y}_{|\mathcal{T}|{+}1}$ and $\mathcal{Y}_{|\mathcal{T}|{+}2}$, respectively.
		The temporary decompression is modeled with a discrete random variable $\ddot{\myrv{T}},\ddot{t}{\in}\ddot{\mathcal{T}}{=}\mathcal{T}\cup\{|\mathcal{T}|{+}1,|\mathcal{T}|{+}2\}$.
		Line \ref{lst:optcluster} optimizes the deterministic mapping $p(t|\ddot{t}){=}\delta(t{-}f_k(\ddot{t}))$ with $f_k:\mathcal{\ddot{T}}{\to}\mathcal{T}$.
		The algorithm restricts merging $y$ into an adjacent cluster $t_1$ or $t_2$. Thus, two mapping options exist $k{\in}\{1,2\}$ with
		\begin{align}
			f_k(\ddot{t}){=}\begin{dcases}
				\ddot{t} & \ddot{t}{\in}\mathcal{T}\\
				t_k & \ddot{t}{=}|\mathcal{T}|{+}1\\
				t_k'& \ddot{t}{=}|\mathcal{T}|{+}2
			\end{dcases}\quad\text{ and $t_k'{=}|\mathcal{T}|{+}1{-}t_k$.}
		\end{align}
		The mutual information loss from merging is
		\begin{align}
			\small
			C_\mylabel{sym}(y,k)=& I(\myrv{X};\ddot{\myrv{T}}|\myrv{S})-I(\myrv{X};\myrv{T}|\myrv{S})\\
			=& \sum_{\ddot{t},s}p(\ddot{t},s)\operatorname{D_{KL}}(p(x|\ddot{t},s)||p(x|f_k(\ddot{t}),s))\\
			=&\sum_s p(s)(C(y,t_k|s){+}C(y',t_k'|s))
			\label{eq:symmergercosts}
		\end{align}
		where the individual merging costs are
		\begin{align}
			\begin{split}
				C(y,t|s){=}& p(\ddot{\myrv{T}}{=}t|s)\operatorname{D_{KL}}\{p(x|\ddot{\myrv{T}}{=}t,s)||p(x|\myrv{T}{=}t,s)\}\\
				+&p(\myrv{Y}{=}y|s)\operatorname{D_{KL}}\{p(x|\myrv{Y}{=}y,s)||p(x|\myrv{T}{=}t,s)\}
			\end{split}
			\label{eq:mergercosts}
		\end{align}
		with $\operatorname{D_{KL}}(p||q){=}\sum_x p(x)\log_2(p(x)/q(x))$.

\section{Performance and Complexity with 5G Codes}
This section investigates the performance and simplifies operations of the proposed memory-assisted decoders.
As in \cite{mohr2024region} we use a 5G-LDPC code with length $8448$, base graph~1 and various code rates\cite{3gpp18}. 
Furthermore, we consider an AWGN channel with BPSK modulation.
All decoders use the initialization schedule described in \cite{mohr2024region} to avoid useless CN updates resulting from processing punctured messages.
If not mentioned otherwise, the remaining schedule follows the flooding scheme with a maximum of 30 decoder iterations. We emphasize that the decoders are designed in an offline-design phase using training data with 10000 transmitted and received code words generated for a design-$E_b/N_0$ which minimizes FER. Thus, all decoder parameters are fixed for all iterations. 

\vspace{-0.2cm}
\subsection{Evolution of Mutual Information}
\begin{figure}[t]
	\centering
	\includestandalone[mode=\mymodestandalone]{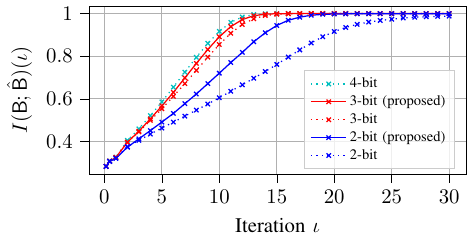} 
	\vspace{-0.25cm}
	\caption{Evolution of mutual information with code rate $1/3$. All results obtained with same design $E_b/N_0{=}1.0$\,dB.}
	\vspace{-0.4cm}
	\label{fig:mievolution}
\end{figure}

\cref{fig:mievolution} depicts the evolution of mutual information between code bit $\B$ and the corresponding hard decision $\hat{\B}$ for every iteration.
The quantized messages are matrix-aligned ($\alpha(\n){=}1$) such that all messages use the same alphabet of reliability levels in one iteration.
For \cref{fig:mievolution} the design process is initialized with the same design-$\EbNo$ for a fair comparison among the configured bit widths.
Particularly under 2-bit decoding, the mutual information gains per iteration are significantly improved with the proposed memory-assistance.
For 3-bit decoding those gains appear smaller. Nevertheless, the proposed 3-bit decoding almost achieves the same performance as the conventional 4-bit decoding.

\vspace{-0.2cm}
\subsection{Boundary Placement for Memory-Assisted Reconstruction}
This section analyzes the placement of quantizer boundaries $\tau_k$ for every iteration to explain the performance gains achieved with the proposed structure. 
Now, all decoders use an individual design-$\EbNo$ so that the mutual information converges to $\I(\myrv{B};\hat{\myrv{B}})(30){\approx} 1-10^{-5}$ after 30 iterations.

\begin{figure}[t]
	\centering
	\includestandalone[mode=\mymodestandalone]{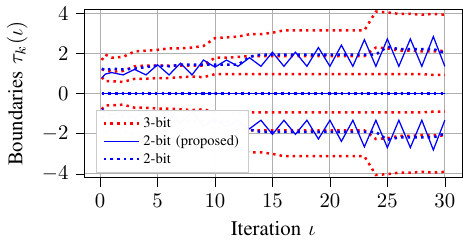} 
	\vspace{-0.25cm}
	\caption{Evolution of quantizer boundary placement each with code rate $1/3$ and individual design $\EbNo=.55$, $.73$ and $1.0$\,dB.
}
\vspace{-0.0cm}
\label{fig:boundaries}
\end{figure}

\cref{fig:boundaries} shows boundary levels which increase for higher iterations as the reliability of messages improves.
One key observation is that boundary magnitudes for the proposed 2-bit decoder show up an alternating rising and falling trend.
Thus, memory-assistance enhances the resolution by using different quantizers in successive iterations.
This behavior shows the effectiveness of the side-information aware IB algorithm: 
A CN message $t$ is a compressed version of the non-quantized LLR $\lc_n$ in \cref{eq:cnu} from the current iteration.
A CN message $\ibs$ is a compressed version of  $\ulc_\n$ from the previous iteration.
The difference $\Delta \lc_n{=}\lc_\n{-}\underline{\ell}^c_n$ is sufficiently small on average, such that $s$ can approximately resolve $\ell^c_n$.
Different boundaries for $\underline{\quant}$ and $\quant$, help to reconstruct $\lc_\n$ from the two coarsely quantized messages $\ibs{=}\underline{Q}(\ulc_n)$ and $t{=}\quant(\lc_\n)$.
The combined resolution approaches the 3-bit quantizer in \cref{fig:boundaries}.%

We remark that the side information $\ibs$ is only actively used during reconstruction as $\phi(t,s)$.
For example, consider a sign-magnitude alphabet sorted by underlying LLR with $t{\in}\{-2,-1,+1,+2\}$. 
For $t{=}{+}2$ the reconstructed LLR $\phi(t{=}{+}2, s)$ is more reliable if the message in memory is matching, e.g., $s{=}{+}2$.
It is less reliable if $s$ does not agree, e.g., $s{=}{-}2$. Thus, $\phi(t{=}{+}2,s{=}{+}2){>}\phi(t{=}{+}2,s{=}{-}2)$.
Hence, additional knowledge of $s$ helps to counteract loss in resolution resulting from coarse quantization with $t{=}Q(\lc_n)$.


\vspace{-.2cm}
\subsection{Complexity Reduction of Memory-Assisted Decoding}\label{sec:reducedcomplexity}
\begin{figure}[t]
	\centering
	\includestandalone[mode=\mymodestandalone]{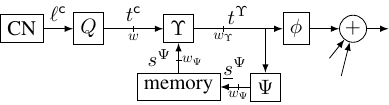} 
	\vspace{-0.2cm}
	\caption{The compression operations $\Upsilon$ and $\Psi$ reduce the complexity for the VN's reconstruction $\phi$ and memory to store side information $s^\Psi$.}
\vspace{-0.0cm}
\label{fig:memawaredesign2}
\end{figure}
The memory-assisted decoder presented so far suffers from additional memory demand and the more expensive reconstruction $\phi(\tc,\scmy)$ shown in \cref{fig:memawaredesign} which requires a lookup table with $2^{2w{-}1}$ entries.
This section proposes a modified setup depicted in \cref{fig:memawaredesign2} with reduced complexity using a new reconstruction $\phi(t^\Upsilon)$ where the input is a message $t^\Upsilon{\in}\mathcal{T}_{w_\Upsilon}$ with $w_\Upsilon{=}w{+}1$.
The message $t^\Upsilon$ is obtained by combining side information $s^\Psi{\in}\mathcal{T}_2$ and a CN message $t^c{\in}\mathcal{T}_w$ using $\Upsilon(s,t){=}\operatorname{sgn}(t)(2|t|{+}s')$, where
\begin{align}
		s'&{=}\Upsilon'(s^\Psi,\operatorname{sgn}(t^c)){=}\begin{dcases}
			1 & \operatorname{sgn}(\tc){=}\operatorname{sgn}(s^\Psi),|s^\Psi|{=}2\\
			0 & \operatorname{otherwise}
		\end{dcases}.
	\label{eq:oldnewcompression}
\end{align}
The side information $\underline{s}^\Psi$ in the next iteration is a compressed version of $t^{\Upsilon}$ where only the sign bit and the most significant magnitude bit are preserved:
\begin{align}
	\underline{s}^\Psi=\Psi(t^{\Upsilon})=\operatorname{sgn}(t^{\Upsilon})\lfloor|t^{\Upsilon}|/2^{w{-}1}\rfloor
	\label{eq:oldcompression}
\end{align}
We remark that the rules \cref{eq:oldnewcompression} and \cref{eq:oldcompression} were discovered by closely inspecting LUTs designed with the IB method realizing $\Upsilon$ and $\Psi$. 
The rules approximate the IB LUTs with high accuracy but much less complexity.
Only two logic gates (XOR, AND) are required to implement \cref{eq:oldnewcompression} and no logic at all is required to implement \cref{eq:oldcompression}.
Moreover, using fixed rules allows the quantizer design for $\tc{=}\quant(\lc)$ to be aware of the subsequent mapping $t^\Upsilon{=}\Upsilon(\tc,s^\Psi)$. 
The design objective becomes $\max_\quant \I(\X;\Tc|\ibS')$ solved with the side information aware IB algorithm using $p(x,\lc,s'){=}\sum_{s^\Psi:s'=\Upsilon'(s^\Psi,\operatorname{sgn}(\lc))}p(x,\lc,s^\Psi)$.

\vspace{-0.15cm}
\subsection{Frame Error Rate Performance using a Flooding Schedule}
\begin{figure*}[t]
	\centering
		\includestandalone[mode=buildmissing]{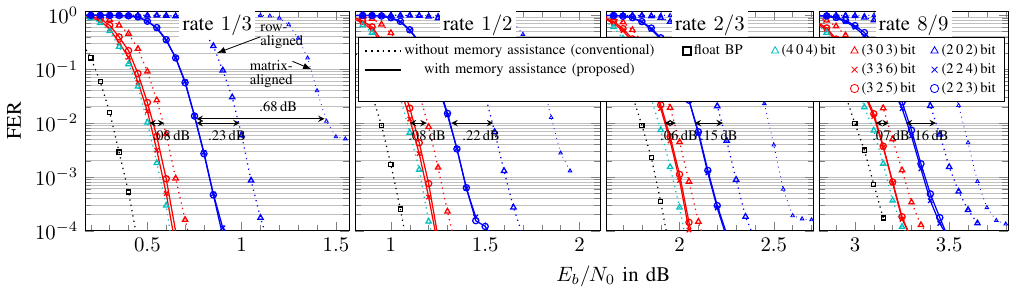} 
	\vspace{-0.4cm}
	\caption{FER performance of memory-assisted (solid lines) and non-assisted (dotted lines) setups for 5G-LDPC decoding (flooding schedule, $\iota_{max}{=}30$).}
	\vspace{-0.5cm}
	\label{fig:fer_5g_rates}
\end{figure*}

The performance using a flooding schedule with different resolutions and code rates is depicted in \cref{fig:fer_5g_rates}.
Unless stated otherwise, all quantized decoders apply CN-aware row-aligned quantization~\cite{mohr2024region}.
The label ($w\,w_\Psi\,w_\Upsilon$) specifies the bit widths: $w$ for exchanged messages, $w_\Psi$ for side information, and $w_\Upsilon$ for reconstruction inputs.
The conventional decoders, labeled ($w\,0\,w$), are designed with analytical tracking of probabilities as in \cite{mohr2024region}. 
The proposed ones from \cref{sec:designmemassistance,sec:reducedcomplexity} labeled ($w\,w\,2w$) and ($w\,2\,(w{+}2)$), respectively, are designed with measured probabilities from training data.
At rate~$1/3$, the proposed $(2\,2\,4)$ memory-assisted decoder improves performance by .23\,dB over the conventional $(2\,0\,2)$ decoder, while the $(3\,3\,6)$ decoder improves by .08\,dB over the $(3\,0\,3)$ decoder.
Relative to a  conventional $(2\,0\,2)$ decoder with matrix alignment, no CN-aware quantization and no memory-assistance, the $(2\,2\,4)$ decoder achieves a gain of .68\,dB.
The $(3\,3\,6)$ memory-assisted decoder is able to slightly outperform a $(4\,0\,4)$ decoder.
The improvements also translate to other code rates $2/3$ to $8/9$. 
Remarkably, the $(w\,2\,w{+}2)$ decoders, using approximations \cref{eq:oldnewcompression} and \cref{eq:oldcompression}, show up almost no performance loss.

\vspace{-0.2cm}
\subsection{Complexity vs. Performance under Layered Decoding}
\begin{figure}
	\centering
	\includegraphics[width=0.99\columnwidth]{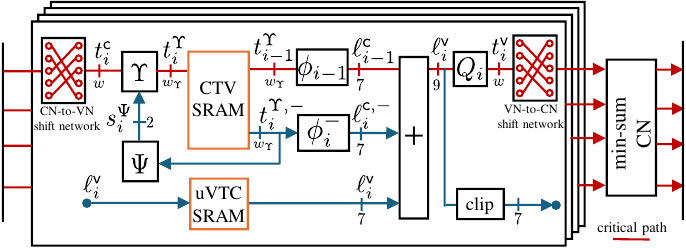}
	\vspace{-0.6cm}
	\caption{A row-parallel decoder architecture with exchange of VN and CN messages (R-VC) from \cite{mohr2024region} extended with memory assistance.}
	\vspace{-0.0cm}
	\label{fig:hardware}
\end{figure}

\cref{fig:hardware} depicts the processing elements for a row-layered architecture from~\cite{mohr2024region}.
Each row $i$ of the base-matrix corresponds to one of $\Nlayersr{=}\lceil22/r\rceil{-}20$ layers depending on rate~$r$.
Thus, $\Z$ rows of the lifted matrix are updated during each pair of steps $(\setUv_k,\setUc_{k+1})$ of the schedule.
We limit the analysis to a partial-parallel processing of each layer with $\Zp{=}32$, i.e., the hardware only processes $\Zp{=}32$ of $\Z{=}384$ rows in parallel to limit the chip area\cite{ren24}.
During $\setUv_k$, a VN update computes $\tv_i{=}\quant_i(\lv_i)$ with $\lv_i{=}\lv_{i{-}1}{+}\lc_{i{-}1}{-}\lcminus_i$.
Lookup tables reconstruct the two LLRs $\lc_{i{-}1}$ and $\lcminus_i$  from the CN message $t^\Upsilon_{i{-}1}$ of the last layer update  and the non-extrinsic CN message from the last iteration $t^{\Upsilon,-}_i$, respectively.
Messages are preserved in static random-access memory (SRAM) between layer updates.
A permutation network performs a cyclic shift of $\Zp{=}32$ messages. A CN update step $\setUc_k$ computes messages for all connected VNs according to~\cref{eq:cnu}.
After an inverse permutation, a CN message $t^c_i$ is merged with the side information $s^{\Psi}_i$.

\cref{table:complexity} evaluates space complexity $\Agr$ by accumulating the logic area to realize the decoder operations of \cref{fig:hardware}, expressed in terms of NAND gate area as in \cite{mohr2024region}.
For the time complexity we count the number of logic gates $\Nd$ on the critical path.
The throughput per iteration is $\TP{\propto}1{/}(\Nd \Nlayersr)$.
The area efficiency is $\AEmy{\propto}\TP/(\iteravg \Agr)$, using the average number of iterations $\iteravg$ until all parity checks are fulfilled.
We normalize results for $\TP$ and $\AEmy$ w.r.t. configuration $(4\,0\,0)$ at rate $1/3$.
Each configuration results in same FER of 0.01 at the given operation-$\EbNo$ by adjusting $\itermax$.
At high rates, $8/9$ to $11/12$, the memory-assisted $(2\,2\,3)$ setting can achieve the best area efficiency.
However, at lower rates, e.g. $1/3$, SRAM area occupies 61\% of the overall area which makes the savings in the remaining processing logic of the decoder less significant.
We remark that relative area for memory can be reduced by increasing parallelism $\Zp$ or using a flooding schedule.

\section{Conclusions}
This letter enhances the reconstruction of coarsely quantized CN messages by utilizing messages from the previous iteration as side-information.
The reconstruction levels depend on the quantization thresholds, whose design is improved with a new side-information-aware IB algorithm.
Further, we proposed a simple rule to combine old and new messages to reduce memory and lookup complexity.
Enabling memory-assistance for a 2-bit (3-bit) flooding decoder achieved SNR gains of up to .23\,dB (.08\,dB).
Finally, we compared the space-time complexity for a row-layered decoder architecture at same FER performance by allowing different maximum iteration count among the configurations. 
Particularly at high code rates where the area used for memory is smaller, our estimates indicate up to a 32\% increase in area efficiency for the 2-bit memory-assisted decoder.

For rate-compatible decoding, a hybrid approach with
memory-assisted high-rate and non-memory-assisted low-rate decoding appears to be useful. In such a setup, high-rate decoding can potentially utilize the additional memory for low-rate decoding to store side information. 
\begin{table}[t]
	\centering
	\caption{Complexity Using the R-VC architecture from \cite{mohr2024region}, Memory-assisted (proposed) settings with bold label}
	\vspace{-0.3cm}
	\label{table:complexity}
	\includestandalone[mode=buildmissing]{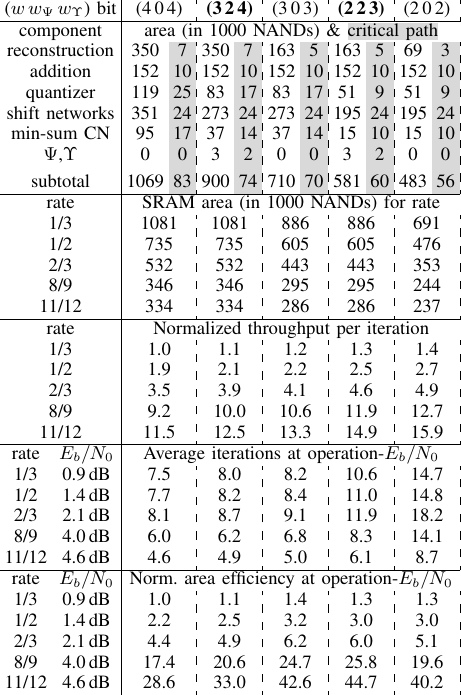}
	\vspace{-0.1cm}
\end{table}

\vspace{-0.31cm}
\bibliographystyle{IEEEtran}
\bibliography{main}
\end{document}